\UseRawInputEncoding
\documentclass[amsmath,amssymb,aps,reprint,superscriptaddress,longbibliography]{revtex4-1}
\usepackage{graphicx}
\usepackage{dcolumn}
\usepackage{bm}

\newcommand\p{\partial}

\newcommand\z{\zeta}

\newcommand\ta{\theta}

\newcommand\ph{\varphi}

\newcommand\eps{\epsilon}

\newcommand\di{\textrm{d}}
\newcommand\ex{\textrm{e}}

\begin{document}

\preprint{APS/123-QED}

\title{Phase field model for cell spreading dynamics  }

\author{Mohammad Abu Hamed}
\affiliation{Department of Mathematics, Technion - Israel Institute of Technology, Haifa 32000, Israel }
\affiliation{Department of Mathematics, The College of Sakhnin - Academic College for Teacher Education, Sakhnin 30810, Israel }

\author{Alexander A. Nepomnyashchy}
\affiliation{Department of Mathematics, Technion - Israel Institute of Technology, Haifa 32000, Israel }

\begin{abstract}
We suggest a 3D phase field model to describe 3D cell spreading on a flat substrate. The model is a simplified version
of a minimal model that was developed in \cite{Winkler+Aranson+Ziebert2019}. Our model couples the order parameter $u$ with 3D polarization (orientation) vector field $\textbf{P}$ of the actin network.
We derive a closed integro-differential equation governing the 3D cell spreading dynamics on a flat substrate, which
includes the normal velocity of the membrane, curvature, volume relaxation rate, a function determined by the
molecular effects of the subcell level, and the adhesion effect. This equation is easily solved numerically. The
results are in agreement with the early fast phase observed experimentally in \cite{Dobereiner2004}. Also we
find agreement with the universal power law \cite{Cuvelier2007} which suggest that cell adhesion or contact area
versus time behave as $\sim t^{1/2}$ in the early stage of cell spreading dynamics, and slow down at the next stages.
\end{abstract}

\maketitle

\section{Introduction}
Understanding the phenomenon of cell spreading has numerous potential applications, which include
designing biomaterials for optimal control of cell
behavior \cite{Ze2018}, insight into cell morphology \cite{Folkman1978}, and
developing efficient methods for gene transfection in biomaterials \cite{Yingjun2019}.

In the last two decades, several models have been developed that describe cell spreading on a flat substrate. Those
models take into account
the elastic \cite{Yuan2015}, \cite{Xiong2010} or visco-elastic properties of the cell and/or the substrate
\cite{Ze2018},
\cite{Nisenholz2014}. In addition, those models describe the dynamics of some subcellular components such as cortical
cytoskeleton, cell nuclear, actin filaments, and microtubules \cite{Fang2016}, \cite{Vernerey2014}. Also, they
consider the mechanical interactions between cell adhesion molecules like cross-membrane protein, molecular clutches,
and the extracellular property of the substrate.     There exist other models that describe cell spreading on
non-flat substrate such as V or Y-shaped micro-patterned substrates \cite{McEvoy2017}. All of the previous models
were compared with and validated by experimental measurements.

Typically, the implementation of computational models needs hard numerical simulations based on finite elements
methods \cite{Vernerey2014},
\cite{Odenthal2013} or minimizing some free energy functionals \cite{Fang2016}; some models include stochastic
effects \cite{McEvoy2017}.

Based on experimental data and measurements, some universality property of cell
spreading have been discovered. Usually early spreading is isotropic. Cell spreading may experience three sequential
phases, basal (cell touches
the substrate), fast continuous spreading (generation of lamellipodial sheet), and periodic local contractile spreading \cite{Dobereiner2004}. These phases obey a  power-law area growth with distinct exponents when we plot cell adhesion area (contact area) versus time. Later the authors in \cite{Cuvelier2007} succeed to explain these power-law
relationships with a relatively simple physical model. They consider energy balance and assume that actin cortex is a
viscous liquid \cite{McGrath2007}.

A minimal computational phase field model of 3D cell crawling on general substrate topography was developed in
\cite{Winkler+Aranson+Ziebert2019}. In the present paper we consider a simplified version of that model. We choose
the substrate to be a flat surface, $z=0$, in order to model the dynamics of cell spreading on the plane. Unlike the
models mentioned above,
our model is simple. We describe the cell spreading dynamics by a single scalar non-local partial differential
equation of the cell interface \eqref{gfd}, which could be solved easily with Wolfram Mathematica program. Our model
is in qualitative agreement with observations at the early fast phase, and the universal power law at the earlier
stages of cell spreading.

The structure of the paper is as follows. In Sec. II we present the minimal 3D phase field model. In Sec. III we
introduce the proper length and time scales of the spreading dynamics. We perform asymptotic analysis and find the
fields at the leading order, and then we use the solvability condition to derive a closed evolutionary nonlocal
equation that describes the cell interface dynamics \eqref{gfd}. We solve this equation numerically via the function
\verb"NDSolve" of Wolfram Mathematica. We reveal the agreement with the universal power law.  Finally, in Sec. IV we
present the conclusions.

\section{Formulation of the problem}
\begin{figure}
  \centering
  \includegraphics[scale=0.3]{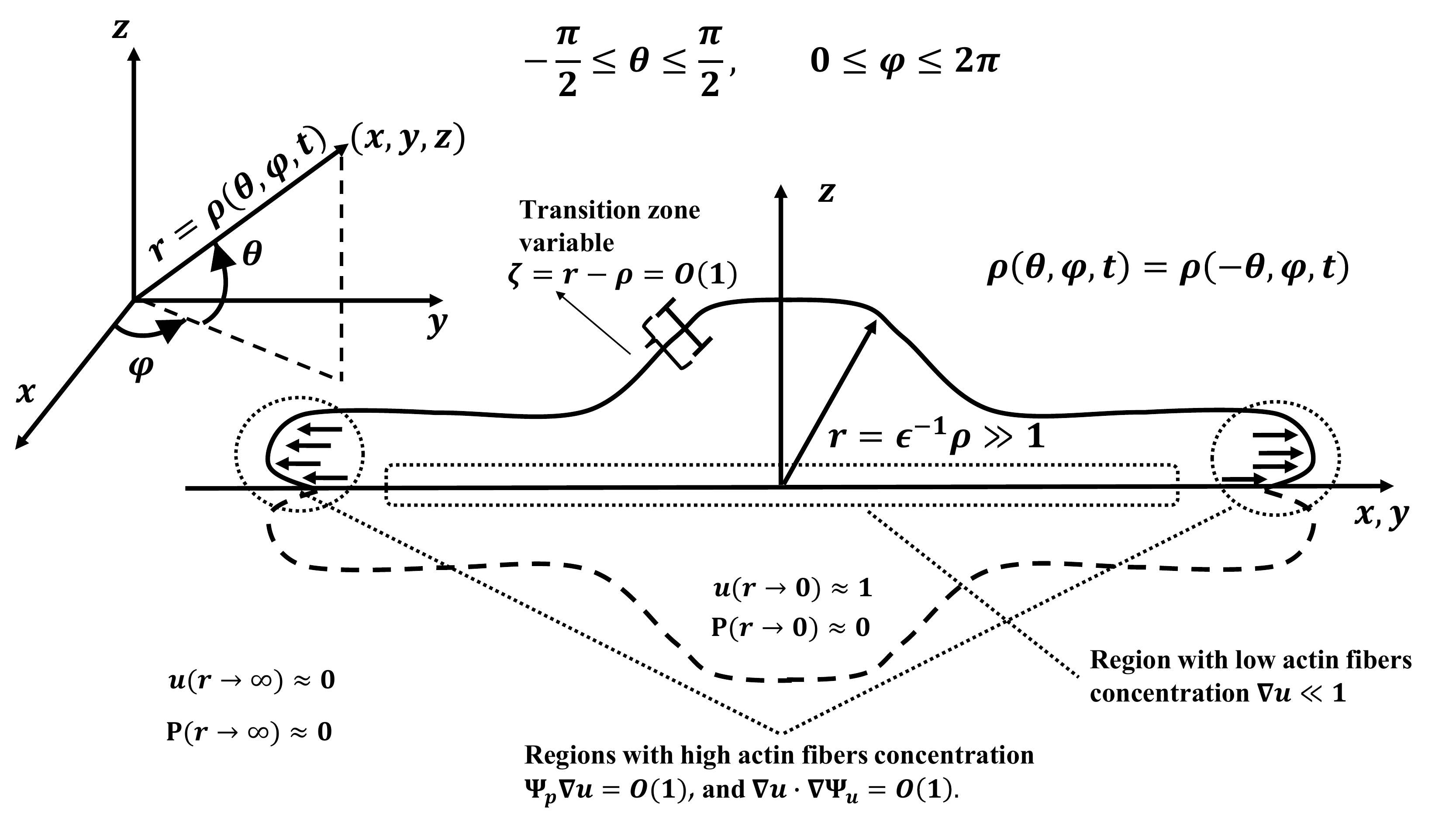}
  \caption{A schematic description of cell spreading dynamics in spherical coordinate system, and the boundary
conditions of the simplified model \eqref{Mod}. The cell is only the shape in the upper region $z\geq0$ while the model \eqref{Mod} is formulated in the hole space with reflection symmetry assumption with respect of the substrate plane $z=0$, \eqref{sys-c}. The thickness of the cell wall (i.e., the width of the transition zone, where $u(r,t)$ is changed from nearly 1 to nearly 0) is $O(1)$, and the cell size is large. Therefore the ratio $\epsilon$ of the thickness of the cell wall to the size of the cell is small. In addition we emphasize the regions where the fields  $\nabla \Psi_u \cdot \nabla u$, and $\Psi_p \nabla u$ give their main contribution and exponentially small otherwise. These are the regions where protrusions developed.   } \label{schematic}
\end{figure}

In order to describe the dynamics of cell located in region $z>0$ and spreading on the flat substrate $z=0$,
see Fig. \ref{schematic}, we extend the problem into the whole space, postulating the reflection symmetry or
antisymmetry of our fields under the transformation $z\to -z$.

Let us consider the following simplified version of the model that was formulated in
\cite{Winkler+Aranson+Ziebert2019}:
\begin{subequations}\label{Mod}
\begin{eqnarray}
&& u_t = D_u \nabla^2 u  -(1-u)(\delta-u)u - \nonumber \\
&& \alpha \nabla u \cdot \textbf{P} - k \nabla \Psi_u \cdot \nabla u ,\label{Mod1} \\
&& \delta = \frac{1}{2} + \mu \delta V -\sigma |\textbf{P}|^2,\ \delta V(t)= \int u \di^3 r - v_0, \label{Mod2}\\
&& \textbf{P}_t = D_p \nabla^2 \textbf{P}  - \tau^{-1}\textbf{P}  - \nonumber \\
&& \beta\Psi_p(z) \left[(1-\nu)\hat{P} \nabla u + \nu\nabla u \right] ,\label{Mod3}\\
&& u(r=0)=1, \quad u(r\rightarrow\infty)=0,\label{Mod4}\\
&& \textbf{P}(r=0)= \textbf{P}(r\rightarrow\infty)=0;\label{Mod5}
\end{eqnarray}
\end{subequations}
see Fig. \ref{schematic}; here $u$ is the order parameter that is close to $1$ inside the cell and $0$ outside,
and $\textbf{P}$ is the three-dimensional polarization vector field representing the actin orientations.
In \eqref{Mod3}, $\hat{P}= \hat{I}-\hat{n}\hat{n}$ is the projection operator onto the local tangential plane, where
 $\hat{n}=\nabla\Psi_p /|\nabla\Psi_p|$ (in our case $\hat{n}=\pm \hat{z}$).
Therefore,
\begin{equation*}\label{}
  \hat{P}\nabla u= \nabla u-\hat{z}\hat{z} \nabla u = u_x \hat{x} + u_y \hat{y}.
\end{equation*}
The parameter $0\leq \nu \leq 1$ model the contribution of actin polarization from the tangential limit $\nu=0$, and
the
isotropic limit $\nu=1$.

The constant parameters of the problem are: $D_u$ is the stiffness of diffuse interface, $D_p$ is the diffusion
coefficient for $\textbf{P}$, $\alpha$ is the coefficient characterizing advection of $u$ by $\textbf{P}$, $\beta$ determines the creation of $\textbf{P}$ at the interface, $\tau^{-1}$ is the inverse time of the degradation of $\textbf{P}$ inside the cell, $v_0$ is twice the overall initial volume of the cell due to the reflection symmetry, $\mu$ is
the stiffness of the volume constraint, and $\sigma$ is the contractility of actin filament bundles. All the parameters listed above are positive.

\begin{figure}
  \centering
  \includegraphics[scale=0.3]{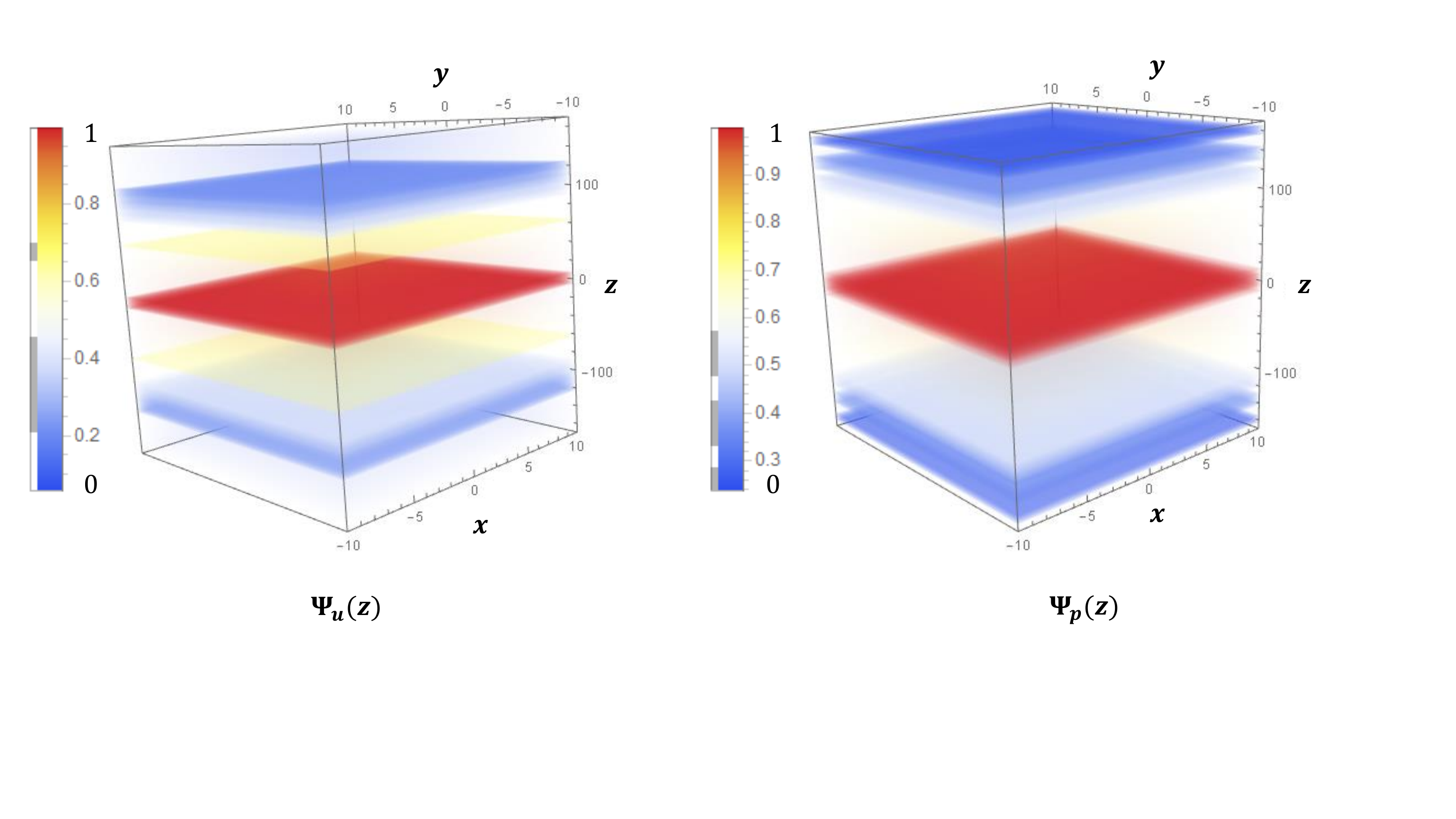}
  \caption{ Plot of the density functions $\Psi_{u,p}$ in \eqref{Psi}. Notice that $\Psi_{u,p}=O(1)$ nearby the substrate $z=0$ (red region) while attenuate far away (blue region) .   The width of the layer where $\Psi_p=O(1)$ is $O(\sqrt{\tau D_p})$. Therefore it is  thicker than layer where $\Psi_u=O(1)$ that is $O(\sqrt{D_u})$.    } \label{density}
\end{figure}

In addition, we make the basic assumption that the ratio $\eps$ of the thickness of the cell wall (i.e., the width
of
the transition zone, where $u$ is changed from nearly 1 to nearly 0) to the characteristic size of the cell is small,
$\eps\ll 1$, see Fig. \ref{schematic}.
Motivated by \cite{Winkler+Aranson+Ziebert2019}, we define the static fields as
\begin{equation}\label{Psi}
  \Psi_u (z)=\ex^{-(\eps z)^2/D_u}, \quad \Psi_p (z)=\ex^{-(\eps z)^2/\l_p}, \ \ l_p=\tau D_p ,
\end{equation}
see Fig. \ref{density}. We choose $l_p>D_u$ to allow more substantial actin inside the cell. The appearance of $\eps$
in the exponents of \eqref{Psi} allows to avoid boundary layer problem complications both in time and space.
The expression $k \nabla \Psi_u \cdot \nabla u$ in \eqref{Mod1} models the adhesion effect of the substrate; $k$
is  adhesion strength parameter. Notice that $\nabla \Psi_u \cdot \nabla u =O(1)$ only in region nearby the substrate and also at the cell boundary or membrane i.e., where protrusion  holds.
  Also the appearance of $\Psi_p \nabla u$  in \eqref{Mod3} allows high actin concentration nearby the flat substrate $z=0$ where protrusions are developed during cell spreading,  and low actin concentration otherwise, see Fig. \ref{schematic}. This scenario is in agreement with experimental studies, see
\cite{Barnhart2011}, \cite{Yuan2015}, and the review paper \cite{Mattila+Lappalainen2008}.

We apply the spherical coordinate system, see Fig. \ref{schematic}, hence $u=u(r,\ta,\ph,t)$,
$\textbf{P}(r,\ta,\ph,t)=p\hat{r}+q\hat{\ta}+w\hat{\ph}$. We
define the iso-surface of the interface as $u(r=\rho(\ta,\ph,t))=1/2$.
    As a result of our definition of the spherical coordinate, we have
\begin{subequations}\label{coor}
\begin{eqnarray}
&&(r,\ta,\ph), \ 0<r<\infty, \ -\frac{\pi}{2}<\ta<\frac{\pi}{2}, \ 0<\ph<2\pi,\\
&& x=r\cos\ta \cos\ph, \ y= r\cos\ta \sin\ph, \ z=r\sin\ta,\\
&& \nabla = \hat{r}\p_r + \hat{\ta}\frac{\p_\ta}{r} + \hat{\ph} \frac{\p_\ph}{r\cos\ta},\\
&& \nabla^2 = \p_{r}^2 + \frac{2\p_r}{r}+ \frac{1}{r^2}\Big( \p_{\ta}^2 -\tan\ta \p_\ta + \frac{\p_\ph^2}{\cos^2 \ta} \Big),\\
&& \nabla^2 \textbf{P} = \hat{r}\nabla^2 p + \hat{\ta}\nabla^2 q + \hat{\ph}\nabla^2 w + O\left(\frac{1}{r^2}\right).
\end{eqnarray}
\end{subequations}

One can calculate,
\begin{eqnarray*}
&&\left[ (1-\nu) \hat{P} + \nu \hat{I} \right]\nabla u =u_x\hat{x} +u_y\hat{y} + \nu u_z\hat{z} = \nabla u+ (\nu-1)u_z\hat{z} \nonumber\\
&&= \left[(1+(\nu-1)\sin^2 \ta)u_r + (\nu-1) \frac{\sin2\ta}{2r} u_\ta \right] \hat{r} +\\
&& \left[(1+(\nu-1)\cos^2 \ta)\frac{u_\ta}{r} + (\nu-1) \frac{\sin2\ta}{2} u_r \right] \hat{\ta} + \frac{u_\ph}{r\cos\ta} \hat{\ph}.
\end{eqnarray*}
Notice that due to the appearance of $\Psi_{u,p}(z)$, the system \eqref{Mod1}-\eqref{Mod5}  does not have any rotationally spherical symmetric solutions. Therefore, we have to look for  general shape solutions.

The reflection symmetry assumption relative to the substrate plane $z=0$, yields the conditions,
\begin{subequations}\label{sys-c}
\begin{eqnarray}
 && u(-\ta)=u(\ta), \quad \rho(-\ta)=\rho(\ta),\\
 && p(-\ta)=p(\ta), \quad w(-\ta)=w(\ta), \quad q(-\ta)=-q(\ta).
\end{eqnarray}
\end{subequations}

\section{Dynamics of general shape interface}

 In order to balance the front dynamics with curvature we impose the following  scaling that describes slow dynamics
of a large-size cell \cite{AbuHamed2020},
\begin{equation}\label{scaling}
  \tilde{t}=\epsilon^2 t, \quad \rho(\ta,\ph,t)=\epsilon^{-1} R(\ta,\ph,t), \quad \eps\ll1 .
\end{equation}
The transition zone variable is defined as
\begin{equation}\label{zeta}
\z=r-\rho(\ta,\ph,t)=O(1).
\end{equation}
Also we define,
\begin{subequations}\label{field-z}
\begin{eqnarray}
 &&  R(\ta,\ph,t) = \tilde{R}(\ta,\ph, \tilde{t}) , \   u(r,\ta,\ph,t) = \tilde{u}(\z,\ta,\ph,\tilde{t}), \\
 &&  \textbf{P}(r,\ta,\ph, t)=\tilde{\textbf{P}}(\z,\ta,\ph, \tilde{t}).
\end{eqnarray}
\end{subequations}
The chain rule yields
\begin{equation}\label{chain1}
  \p_t = -\eps  \tilde{R}_{\tilde{t}}\p_\z + \eps^2 \p_{\tilde{t}}, \quad \p_r = \p_\z.
\end{equation}
later on we drop the tildes.
 It holds that
 \begin{subequations}\label{chain2}
 \begin{eqnarray}
 && \frac{1}{r} = \frac{\epsilon}{R} - \frac{\epsilon^2 \z}{R^2 } +..., \\
 && \Psi_p(z)=\ex^{-(\eps z)^2 /l_p}=\ex^{- \eps^2 r^2 \sin^2 \ta/l_p}=\nonumber\\
 && \ex^{-\eps^2(\zeta+\epsilon^{-1}R)^2 \sin^2 \ta /l_p}\sim \ex^{-R^2 \sin^2 \ta /l_p},\\
 && \nabla \Psi_u \sim \frac{-\eps R}{D_u}\ex^{-R^2 \sin^2 \ta /D_u}\left( 2\sin^2 \ta \hat{r} + \sin2\ta \hat{\ta} \right).
\end{eqnarray}
\end{subequations}
In addition one can calculate,
\begin{subequations}
\begin{eqnarray}
&& \p_\ta = -\eps^{-1} R_\ta \p_\z + \p_\ta, \ \p_\ph = -\eps^{-1} R_\ph \p_\z + \p_\ph,\\
&& \p_\ta^2  = \eps^{-2} R_\ta^2 \p_{\z}^2 - \eps^{-1}( R_{\ta\ta}\p_\z +2R_\ta \p_{\z\ta}^2 ) + \p_{\ta}^2,\\
&& \p_\ph^2  = \eps^{-2} R_\ph^2 \p_{\z}^2 - \eps^{-1}( R_{\ph\ph}\p_\z +2R_\ph \p_{\z\ph}^2 ) + \p_{\ph}^2,\\
&& \frac{\p_\ta}{r} = -\frac{R_\ta}{R} \p_\z +O(\eps),\  \frac{\p_\ta}{r^2} = -\eps \frac{R_\ta}{R^2} \p_\z +O(\eps^2),\\
&& \frac{\p_\ph}{r} = -\frac{R_\ph}{R} \p_\z +O(\eps),\  \frac{\p_\ph}{r^2} = -\eps \frac{R_\ph}{R^2} \p_\z +O(\eps^2),\\
&& \nabla^2 u = \left( 1 + \frac{R_\ta^2}{R^2} +\frac{R_\ph^2}{R^2 \cos^2 \ta} \right) u_{\z\z} + \nonumber\\
&& \eps \Bigg[   \left( \frac{2}{R} + \frac{R_\ta}{R^2}\tan\ta - \frac{R_{\ta\ta}}{R^2} - \frac{R_{\ph\ph}}{R^2 \cos^2 \ta }\right)u_\z \nonumber\\
&&  -\frac{2}{R^2}\left( R_\ta u_{\z\ta} + \frac{R_\ph}{\cos^2 \ta}u_{\z\ph} \right) \nonumber \\
&& -\frac{2\z}{R^3} \left(R_\ta^2 + \frac{R_\ph^2}{\cos^2 \ta} \right) u_{\z\z}  \Bigg]+O(\eps^2). \label{LO}
\end{eqnarray}
\end{subequations}
We can approximate the  nonlocality in (\ref{Mod2}) as follows,
 \begin{equation}\label{nonloc}
   \int u \di^3 r \sim   \frac{\eps^{-3}}{3} \int_{0}^{2\pi} \di \ph \int_{-\pi/2}^{\pi/2} R^3 (\ta,\ph,t) \cos\ta  \di \ta .
 \end{equation}
Consider the following scaling of the model parameters
\begin{equation}\label{scal1}
 \alpha = \eps A, \   \frac{4\pi\mu}{3}\eps^{-3} = \eps M, \ \sigma = \eps S, \ v_0=\eps^{-3}V_0, \  k=O(1).
\end{equation}
  Let us introduce the expansions
 \begin{equation}\label{expan}
  u = u_0 + \eps  u_1+..., \quad p = p_0 + \eps p_1 +...
\end{equation}
We define the auxiliary function , 
\begin{equation*}\label{}
   \Lambda(\ta,\ph,t)= \left( 1+ \frac{R_\ta^2}{R^2} + \frac{R_\ph^2}{R^2 \cos^2 \ta} \right)^{-1/2} ,
 \end{equation*}
 and the function,
 \begin{eqnarray*}
   &&\Phi(\tau,D_u , D_p, \z  ) =\label{Phi(z)}\\
  &&\frac{1}{8}\sqrt{\frac{\tau}{2 D_u D_p}} \int_{-\infty}^{\infty} \ex^{-|s|/\sqrt{\tau D_p}} \cosh^{-2} \left( \frac{s - \z}{\sqrt{8D_u}} \right) \di s, \nonumber
 \end{eqnarray*}
 that are basic for our next analysis.

We substitute the length, time \eqref{scaling}, and the parameters scaling \eqref{scal1}  into system \eqref{Mod}. We
write the system \eqref{Mod} in the transition zone variable \eqref{zeta}-\eqref{field-z} via the expansions and the chain rules \eqref{chain1}-\eqref{nonloc}.  We substitute the asymptotic expansions \eqref{expan} and finally we collect terms of the same order.

Consequently we obtain at the leading order the following system,
 \begin{eqnarray*}\label{}
&& D_u \Lambda^{-2} u_{0\z\z} = (1-u_0)(\frac{1}{2}-u_0)u_0,\\
&& D_p \Lambda^{-2} p_{0\z\z} - \tau^{-1} p_0 =\\
&& \beta \ex^{-R^2 \sin^2 \ta /l_p}\left[1+(\nu-1)\sin^2 \ta - (\nu-1)\frac{\sin2\ta}{2} \frac{R_\ta}{R}  \right] u_{0\z},\nonumber\\
&& D_p \Lambda^{-2} q_{0\z\z} - \tau^{-1} q_0 =\\
&& -\beta\ex^{-R^2 \sin^2 \ta /l_p } \left[(1+(\nu-1)\cos^2 \ta) \frac{R_\ta}{R} - (\nu-1)\frac{\sin2\ta}{2} \right] u_{0\z},\nonumber\\
&& D_p \Lambda^{-2} w_{0\z\z} - \tau^{-1} w_0 = -\beta \ex^{-R^2 \sin^2 \ta /l_p} \frac{R_\ph}{R\cos\ta} u_{0\z}.
\end{eqnarray*}
Following the Ginzburg-Landau theory and Fourier transform method the  solution of this system is given by,
\begin{subequations}
 \begin{eqnarray*}\label{}
&& u_0 (\z) = \frac{1}{2} \left[ 1-\tanh\left(\frac{\Lambda\z}{\sqrt{8D_u}}\right) \right], \label{l1}\\
&& p_0 (\z)  = \beta \lambda_p \Lambda\Phi(\Lambda\z),\label{l2} \\
&&\lambda_q = \ex^{-R^2 \sin^2 \ta /l_p }\left[1+(\nu-1)\sin^2 \ta - (\nu-1)\frac{\sin2\ta}{2} \frac{R_\ta}{R}  \right],  \nonumber\\
&& q_0(\z) =-\beta \lambda_q  \Lambda\Phi(\Lambda\z),\label{l3}\\
&& \lambda_q =  \ex^{-R^2 \sin^2 \ta /l_p} \left[(1+(\nu-1)\cos^2 \ta) \frac{R_\ta}{R} - (\nu-1)\frac{\sin2\ta}{2} \right],\nonumber\\
&& w_0(\z) = -\beta \lambda_w \Lambda \Phi(\Lambda\z),\label{l4}\\
&&  \lambda_w = \ex^{-R^2 \sin^2 \ta /l_p} \frac{R_\ph}{R\cos\ta}.\nonumber
\end{eqnarray*}
\end{subequations}
Notice that these results satisfy the symmetry conditions \eqref{sys-c}, if $R(-\theta)=R(\theta)$.

The equation for the correction term $u_1$ at the order $O(\eps)$ have the form,
\begin{eqnarray}\label{}
&& L[u_1]=\textrm{RHS},\label{curef}\\
&&L =  D_u \Lambda^{-2} \p_\z^2  - \left(\frac{1}{2} -3 u_0 + 3u_0^2 \right)\hat{I},\nonumber\\
&& \textrm{RHS}=  - R_t u_{0\z}-\nonumber\\
&&D_u \Bigg[ \left( \frac{2}{R} + \frac{R_\ta}{R^2}\tan\ta - \frac{R_{\ta\ta}}{R^2} - \frac{R_{\ph\ph}}{R^2 \cos^2 \ta }\right)u_{0\z} \nonumber\\
&&  -\frac{2}{R^2}\left( R_\ta u_{0\z\ta} + \frac{R_\ph}{\cos^2 \ta}u_{0\z\ph} \right) \nonumber \\
&& -\frac{2\z}{R^3} \left(R_\ta^2 + \frac{R_\ph^2}{\cos^2 \ta} \right) u_{0\z\z}  \Bigg]\nonumber\\
&& + A  \left( p_0 -\frac{R_\ta}{R} q_0 - \frac{R_\ph}{R\cos\ta}w_0  \right)u_{0\z} \nonumber\\
&& -\frac{k}{D_u} R \cdot \ex^{-R^2 \sin^2 \ta /D_u}\left( 2\sin^2 \ta  - \sin(2\ta) \frac{R_\ta}{R} \right)u_{0\z} +\nonumber\\
&& (1-u_0)u_0 \left\{  \tilde{V}(t) -  S(p_0^2 + q_0^2 + w_0^2)  \right\}, \nonumber
\end{eqnarray}
 where the volume variation have the form,
 \begin{equation*}
  \tilde{V} (t)= M\left[ \frac{1}{4\pi}  \int_{0}^{2\pi} \di \ph \int_{-\pi/2}^{\pi/2} R^3 (\ta,\ph,t) \cos\ta  \di \ta - \frac{3}{4\pi}V_0 \right].
 \end{equation*}

 We apply the solvability condition, which is the orthogonality of the  right-hand side (RHS) of equation \eqref{curef} to the solution $u_{0\z}$ of the homogenous equation $L[u]=0$ of \eqref{curef} i.e.,
\begin{equation*}\label{}
\int_{-\infty}^{\infty}\textrm{RHS} (\z)\cdot u_{0\z} (\z) \di \z =0.
\end{equation*}
We therefore obtain a closed equation governing the interface dynamics $R(\ta,\ph,t)$,
  \begin{equation}\label{gfd}
   a \Lambda R_t = - 2a D_u \mathcal{H} - \tilde{V} + \Omega - N,
 \end{equation}
 where
 \begin{equation}\label{curv}
 \mathcal{H} =\frac{1}{2} \nabla\cdot \hat{n} = \frac{1}{2} \nabla\cdot \left( \frac{\nabla(r-R)}{|\nabla(r-R)|} \right)
 \end{equation}
 is the mean local curvature of the surface $r= R(\ta,\ph,t)$, see Appendix \ref{A}, and
 \begin{eqnarray*}
&& \Omega(\ta,\ph,t) = 6\beta A \Omega_1 \Lambda^2 \left(\lambda_p + \frac{R_\ta}{R}\lambda_q + \frac{R_\ph}{R\cos\ta}\lambda_w \right)+\nonumber\\
&& 6\beta^2 S \Omega_2 \Lambda^2  \left(\lambda_p^2 + \lambda_q^2 + \lambda_w^2 \right),\\
&& \Omega_1 (\tau ,D_u , D_p  ) = \int_{-\infty}^{\infty} \Phi(\xi) \bar{u}_{0\xi}^2(\xi) \di \xi, \\
&& \Omega_2 (\tau, D_u , D_p  ) = \nonumber \\
&&\int_{-\infty}^{\infty} \Phi^2 (\xi) (\bar{u}_0(\xi)-1) \bar{u}_0(\xi) \bar{u}_{0\xi}(\xi) \di \xi>0,\\
 \end{eqnarray*}
 where
 \begin{equation*}\label{}
\bar{u}_0 (\xi) = \frac{1}{2} \left[ 1-\tanh\left(\frac{\xi}{\sqrt{8D_u}}\right) \right].
 \end{equation*}
 The adhesion effect is implemented by
 \begin{equation*}\label{}
   N= \frac{a k}{D_u} \Lambda R \cdot \ex^{-R^2 \sin^2 \ta /D_u}\left( 2\sin^2 \ta  - \sin(2\ta) \frac{R_\ta}{R} \right)
 \end{equation*}

For more details  and explanations about arguments for the  derivation of the governing equation \eqref{gfd} see
\cite{AbuHamed2016}, \cite{AbuHamed2020}, and \cite{AbuHamed2021}.

Equation \eqref{gfd} comes in conjunction with the Neumann boundary conditions
 \begin{equation}\label{BC}
   R_\ta(\ta=0)=R_\ta(\ta=\pi/2)=0,
 \end{equation}
and some initial interface $R(t=0)$.

In order to model cell spreading, we may consider the axi-symmetric case $R_\ph=0$, since according to experimental
observation, the onset of cell spreading is isotropic \cite{Dobereiner2004}. As for the initial interface shape, we
take
the truncated sphere with radius $R_0$ and center $(0,0,\eta)$, see Fig. \ref{t-seq}(a).
\begin{equation}\label{i-S}
R(t=0)=\eta\sin\ta + \sqrt{R_0^2-\eta^2 \cos^2 \ta},  \ \ 0\leq \eta \leq 1.
 \end{equation}
Then the initial normalized volume in this case is given by
\begin{equation*}\label{}
V_0=\int_{0}^{\pi/2}R(t=0)^3\cos\ta\di \ta,
 \end{equation*}
which is twice the volume of the truncated sphere.

In Fig. \ref{t-seq}(a)-(f) we present the sequence of plots that show the results of the numerical simulation of
the interface $R(\ta,\ph,t)$ according to
equation \eqref{gfd} and boundary condition \eqref{BC}. We use the function \verb"NDSolve" of Wolfram Mathematica.
We plot $R(\ta,\ph,t)$ only in the upper region
$0\leq\ta\leq\pi/2$, the plot in the lower region is only a mirror reflection of the upper surface due to our symmetry assumption \eqref{sys-c}.
 Following experimental scenarios where the spherical-like cell almost touch the substrate we may take the parameters of our initial interface as $\eta=0.95$, and $R_0=1$, see Fig. \ref{t-seq}(a). This simulation as we see describes cell spreading. We begin from almost full sphere and end up with ellipsoid-like shape which is the steady state solution of the system \eqref{gfd}, \eqref{BC}, and \eqref{i-S}.  In Fig. \ref{t-seq}(g) we display the cell hight $R(\pi/2,t)$ which decreases from almost 2 to 1.21, while in Fig. \ref{t-seq}(h) we display the cell contact area (radius) $R(0,t)$ which increases to 2.

 In addition Fig. \ref{t-seq}(g)-(h) display the fast spreading phenomena at the beginning  of cell spreading in
agreement with the continuous spreading fast phase that was observed experimentally in \cite{Dobereiner2004}.
Also,in
Fig \ref{t-seq}(i) we consider the Log-Log plot of the cell radius versus time, also we plot the piecewise function
that connect two function of the form $b_1 t^{1/2}$, and $b_2 t^{1/4}$, for a proper choice of the parameters
$b_{1,2}$, and for the connecting point. We notice the agreement with the universal power law \cite{Cuvelier2007} that suggest that cell adhesion or contact area versus time behave as $\sim t^{1/2}$ in the early state of cell spreading dynamics, and slow down in the next states. The plot of the slope $\sim t^{1/4}$ in Fig. \ref{t-seq}(i) is only to emphasize the slowing down of the next phase.

  In Fig. \ref{t-seq1} we perform similar analysis where we choose $D_u=0.5$, and $D_p=0.02$ while the other parameters remain as those of  Fig. \ref{t-seq}. Notice that in this case we have $\tau D_p < D_u$, unlike the previous case of Fig. \ref{t-seq}

\section{Conclusion}
We utilize a simplified version of minimal 3D phase field model that was developed in
\cite{Winkler+Aranson+Ziebert2019}, in order to model cell spreading dynamics on a flat substrate. The model
\eqref{Mod} couples the order parameter $u$ with 3D polarization (orientation) vector field $\textbf{P}$ of the actin
network. The model is formulated in the whole space but with appropriate symmetry conditions with respect to
transformation $z\rightarrow -z$, \eqref{sys-c}.

After we introduce the proper time and length scale and perform asymptotic expansion, we solve equations for the
fields at
the leading order. As a result of the solvability condition we derive a closed integro-differential equation \eqref{gfd} governing the 3D cell spreading dynamics, which includes the normal velocity $\omega_n=\Lambda R_t$ of the membrane, curvature $\mathcal{H}$, volume relaxation rate $\tilde{V}$, a function $\Omega(t)$ determined by the molecular effects of the subcell level, and the adhesion effect $N$.

Excluding the adhesion effect  this result is similar to the 2D case which describe the onset of 2D cell dynamics on flat substrate \cite{AbuHamed2020} and 3D case that describe the onset of 3D cell motility immersed in 3D extracellular matrices  \cite{AbuHamed2021}.

The equation governing the interface or membrane dynamics during spreading may be presented in the form:
 \begin{equation*}\label{}
  \omega_n = - 2 D_u \mathcal{H} - \tilde{V} + \Omega - N,
 \end{equation*}
   after we put the proper scaling transformation, $t\rightarrow a^2 t $ and $R(t)\rightarrow aR(t) $ in \eqref{gfd}.
This equation is easily solved numerically via the function \verb"NDSolve" of Wolfram Mathematica. The simulation
present cell spreading with significant hight decreasing and radius increasing of the initial truncated spheres, see Fig \ref{t-seq}, Fig.  \ref{t-seq1}

   These results are in agreement with the early fast phase that was observed experimentally in
\cite{Dobereiner2004}.
Surprisingly, the result are in qualitative agreement with universal power law which suggest that adhesion or contact
area versus time behave as $\sim t^{1/2}$ in the early state of cell spreading dynamics ,and then it slow down. The
appearance of the slope $\sim t^{1/4}$ in Fig. \ref{t-seq}(i) is only to emphasize the slowing down of the later
phase. This is a surprising result since in our phase field model we did not assume any viscosity property of the
cell membrane as it is assumed in \cite{Cuvelier2007}.

\appendix\section{}\label{A}
Here we give an explicit expression for the mean curvature of a surface given in spherical coordinate description
$r=R(\ta,\ph,t)$, see Fig. \ref{schematic}. Following the definition \eqref{curv},\eqref{coor},
one can calculate,
\begin{eqnarray}\label{}
  &&  \nabla \cdot \hat{n} = \Lambda \Bigg\{ \frac{2}{R} - \frac{R_{\ta\ta}}{R^2} + \frac{R_\ta \tan\ta}{R^2} - \frac{R_{\ph\ph}}{R^2 \cos^2\ta} \Bigg\}\nonumber\\
  && +\Lambda^3 \Bigg\{ \frac{1}{R^3}\left( R_\ta^2 + \frac{R_\ph^2}{\cos^2 \ta} \right)+ \frac{R_\ta^2 R_{\ta\ta}}{R^4} +\frac{2R_\ta R_\ph R_{\ta\ph}}{R^4 \cos^2 \ta}\nonumber\\
  &&+ \frac{R_\ph^2}{R^4 \cos^4 \ta } \left( \frac{1}{2} R_\ta  \sin2\ta +  R_{\ph\ph}  \right)    \Bigg\}
 \end{eqnarray}

\begin{figure}
  \centering
  \includegraphics[scale=0.25]{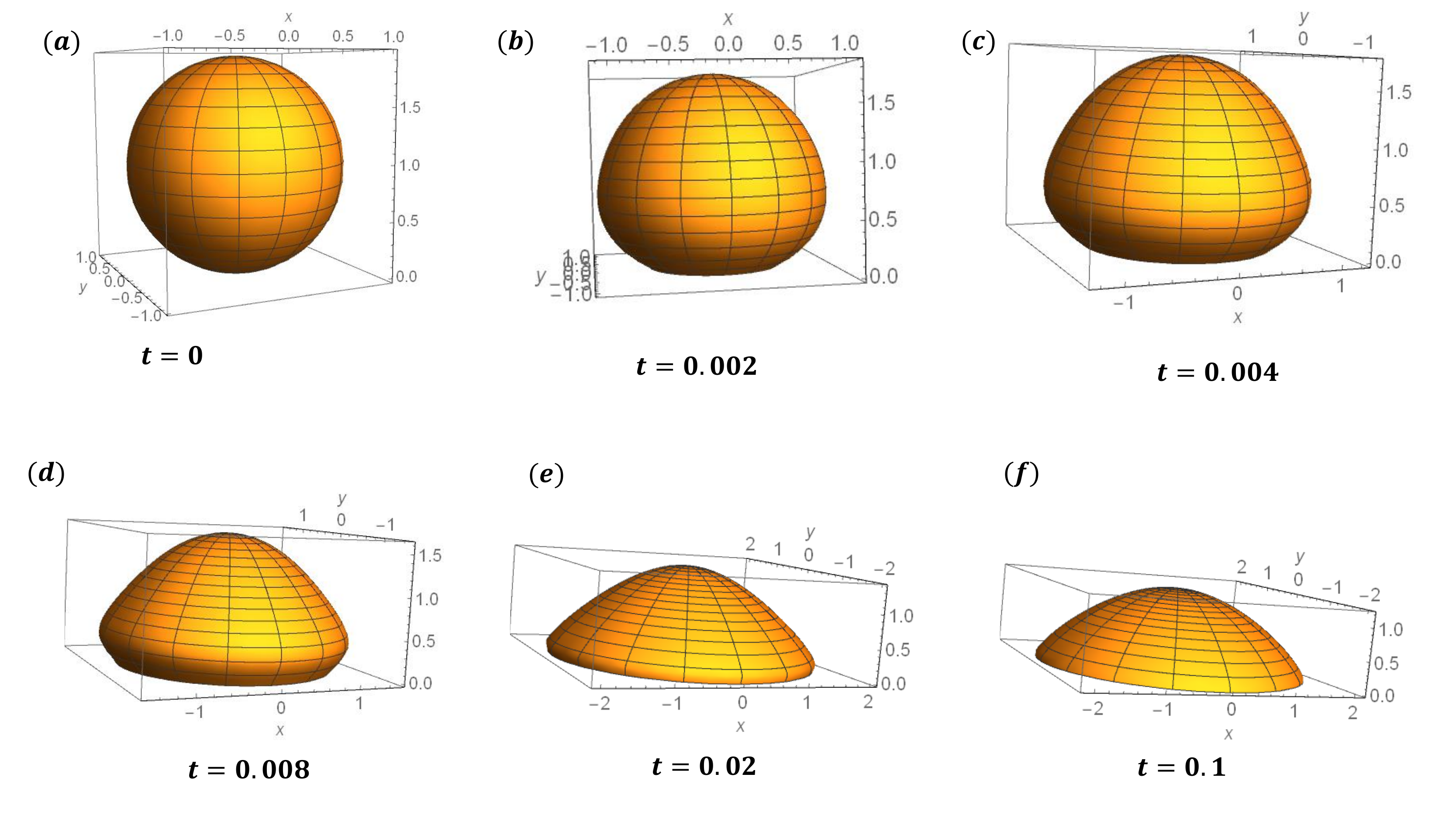}
  \includegraphics[scale=0.25]{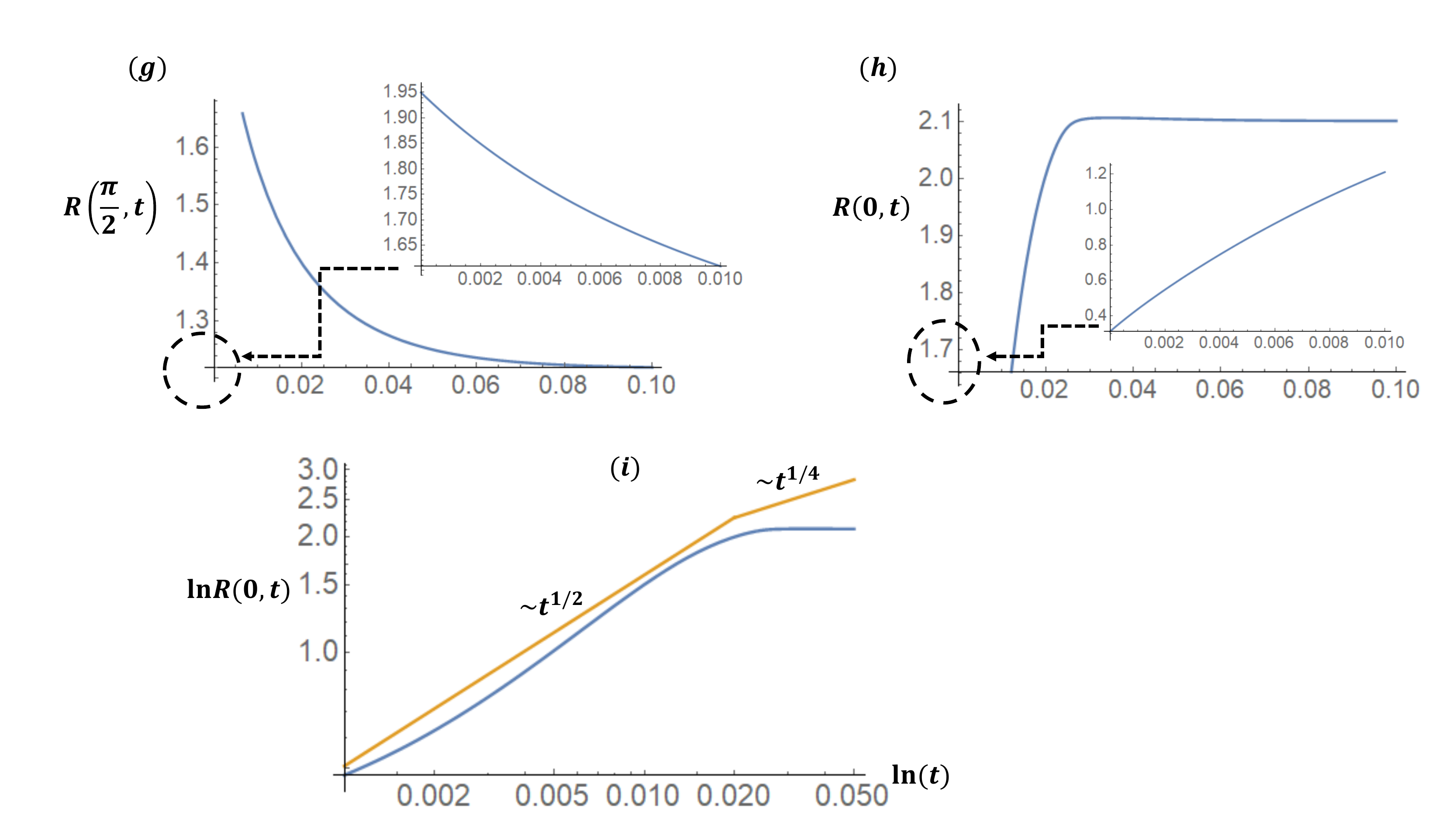}
  \caption{(a)-(f) Time sequence of the simulation of the axisymmetric case of equation \eqref{gfd} with initial
upwards shifted sphere \eqref{i-S} with radius $R_0=1$ and center $(0,0,0.95)$.  We employ the following value of
parameters $\beta=5, A=1, \tau=10,D_u=1,D_p=0.2,M=8,S=2,\nu=0.5, k=15$.  (g) is the plot of the  ellipsoid like hight
$R(\pi/2,t)$, (h) is the plot of the  ellipsoid like radius  $R(0,t)$, both in the time interval $0\leq t \leq 0.1$.
Notice the stationary ellipsoid-like has hight 1.21 and radii 2. (i) Log-Log plot of the cell radius $R(0,t)$ verses
time $t$. Also we plot the piecewise function that connect two function of the form $b_1 t^{1/2}$, and $b_2 t^{1/4}$,
for a proper choice of the parameters $b_{1,2}$, and for the connecting point. Notice the qualitative agreement with
the universal power law in the initial fast phase and next the slower phase. The appearance of the slope $\sim
t^{1/4}$ in (i), is shown only to emphasize the slowing down of the later phase. } \label{t-seq}
\end{figure}

\begin{figure}
  \centering
  \includegraphics[scale=0.25]{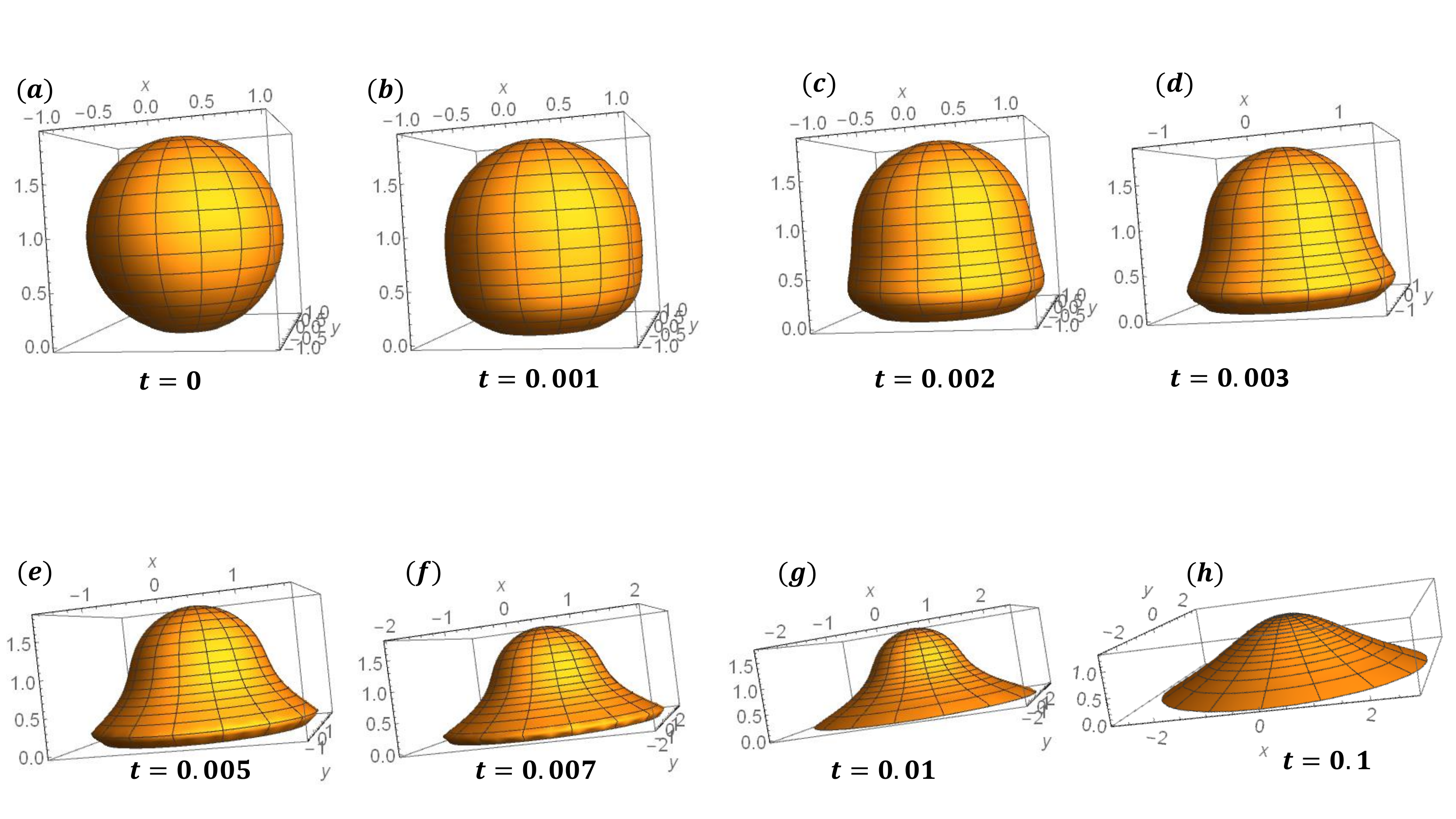}
   \includegraphics[scale=0.25]{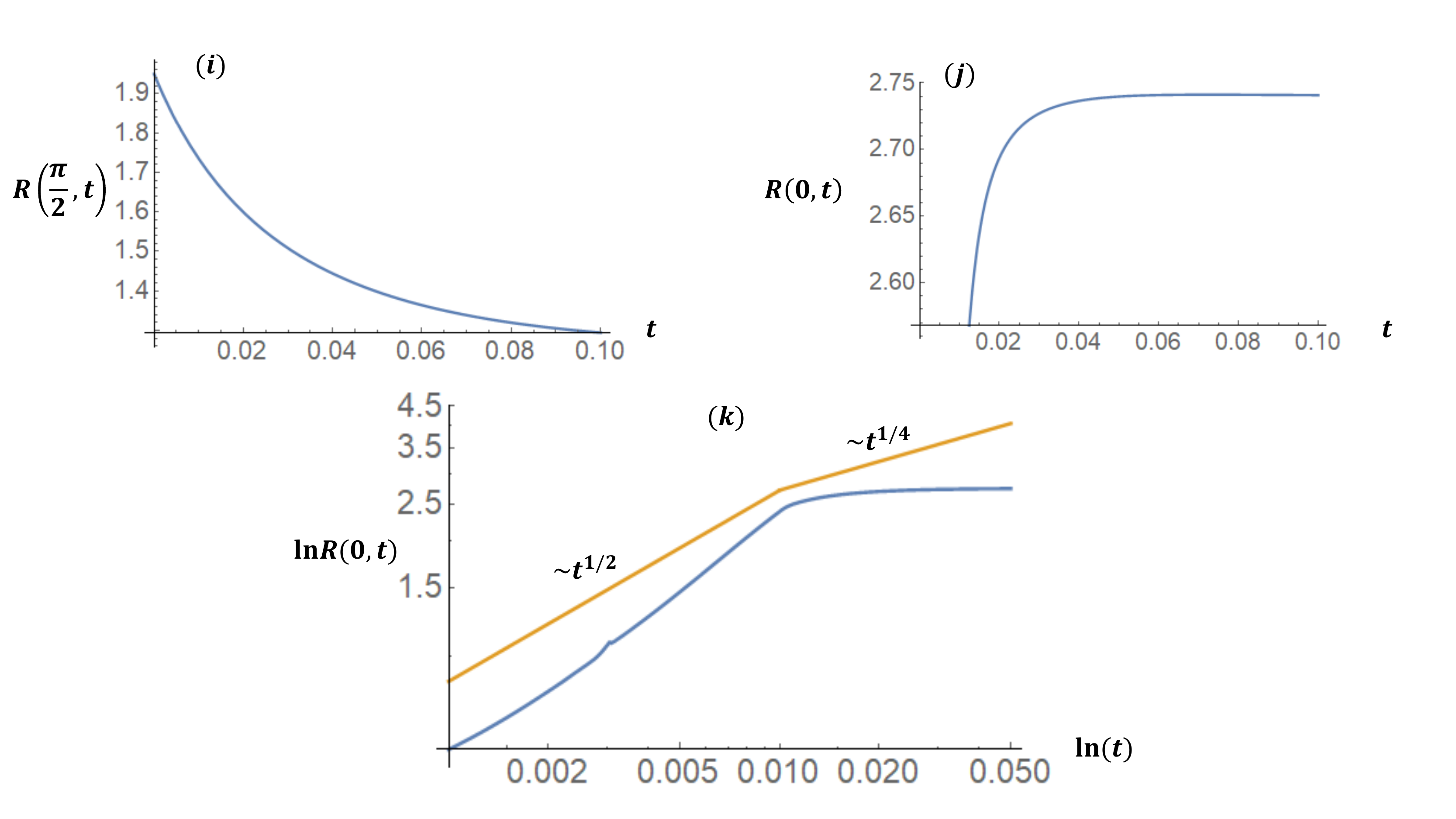}
  \caption{(a)-(h) Time sequence of the simulation of the axi symmetric case of equation \eqref{gfd} .  We employ following value of parameters $D_u=0.5,D_p=0.02$, the other parameters and the initial sphere are as in Fig \ref{t-seq}.  (i) is the plot of the  ellipsoid like hight  $R(\pi/2,t)$, (j) is the plot of the  ellipsoid like radius  $R(0,t)$, both in the time interval $0\leq t \leq 0.1$. Notice the stationary ellipsoid-like has hight 1.3 and radii 2.74. Notice that here we have $\tau D_p < D_u$, unlike the previous case of Fig. \ref{t-seq}. (k) Log-Log plot as that of Fig. \ref{t-seq}.      } \label{t-seq1}
\end{figure}

\clearpage
\bibliography{HSCD2.4a.bbl}{}

\end{document}